\documentclass[aps,pre,preprint,groupedaddress,showpacs]{revtex4-2}

\usepackage{graphicx}
\usepackage{amssymb}
\usepackage{color} 
\usepackage{epstopdf}
\usepackage{soul}
\newcommand{\corr}[1]{\textcolor{black}{#1}} 

\begin{document}


\title{Cratering by the oblique impact of a spinning projectile\\
	\textnormal{Accepted manuscript for Physical Review E, 113, 035412, (2026), DOI: 10.1103/zn6h-97dp, https://doi.org/10.1103/zn6h-97dp}}



\author{Douglas D. Carvalho}
\author{Erick M. Franklin}
 \email{erick.franklin@unicamp.br}
 \thanks{Corresponding author}
\affiliation{%
	Faculdade de Engenharia Mec\^anica, Universidade Estadual de Campinas (UNICAMP)\\
	Rua Mendeleyev, 200, Campinas, SP, Brazil\\
}%


\date{\today}

\begin{abstract}
We investigate the roles of spin and packing fraction on the dynamics of cratering when a solid projectile impacts a granular bed at different incident angles. For that, we carried out DEM (discrete element method) computations in which we varied the magnitude and direction of the projectile spin, the impact velocity, the bed packing fraction, and the incident angle. For a given incident velocity, we found that the projectile can rebound for small angles, or be completely or partially buried for larger angles, and that when buried it can sometimes migrate large horizontal distances depending on the incident angle. We also found that increasing the packing fraction strengthens rebounds, and that the initial spin, depending on its direction and orientation, induces rebound, burying, or transverse deviations. The crater morphology also changes with the varying parameters, acquiring circular, elliptical, goutte-like, tadpole-like, and transitional shapes, correlating well with the projectile behavior. Finally, we propose diagrams organizing and classifying the dynamics observed. Our results shed new light on the different shapes of craters found in nature and the fate of the impacting material.  
	
\end{abstract}


\maketitle


\section{INTRODUCTION}
\label{sec:intro}

The formation of craters by the impact of a projectile is frequent in both nature and human activities, and occurs in a large range of scales: from the low energy impacts (as low as 10$^{-7}$ J) of seeds falling on the ground to the high energy impacts (from 10$^{16}$ J on) of asteroids colliding with planets and moons \cite{Melosh, Holsapple}. Understanding the dynamics of cratering is, thus, important for agriculture and forest conservation (spreading, dropping, and burying of seeds), mining (predicting the final positions of rare minerals brought to Earth by falling asteroids), and geophysics in general (understanding land-surface processes and the existing landscapes of planets and moons), for example.

\begin{figure}[h!]
	\begin{center}
		\includegraphics[width=0.7\linewidth]{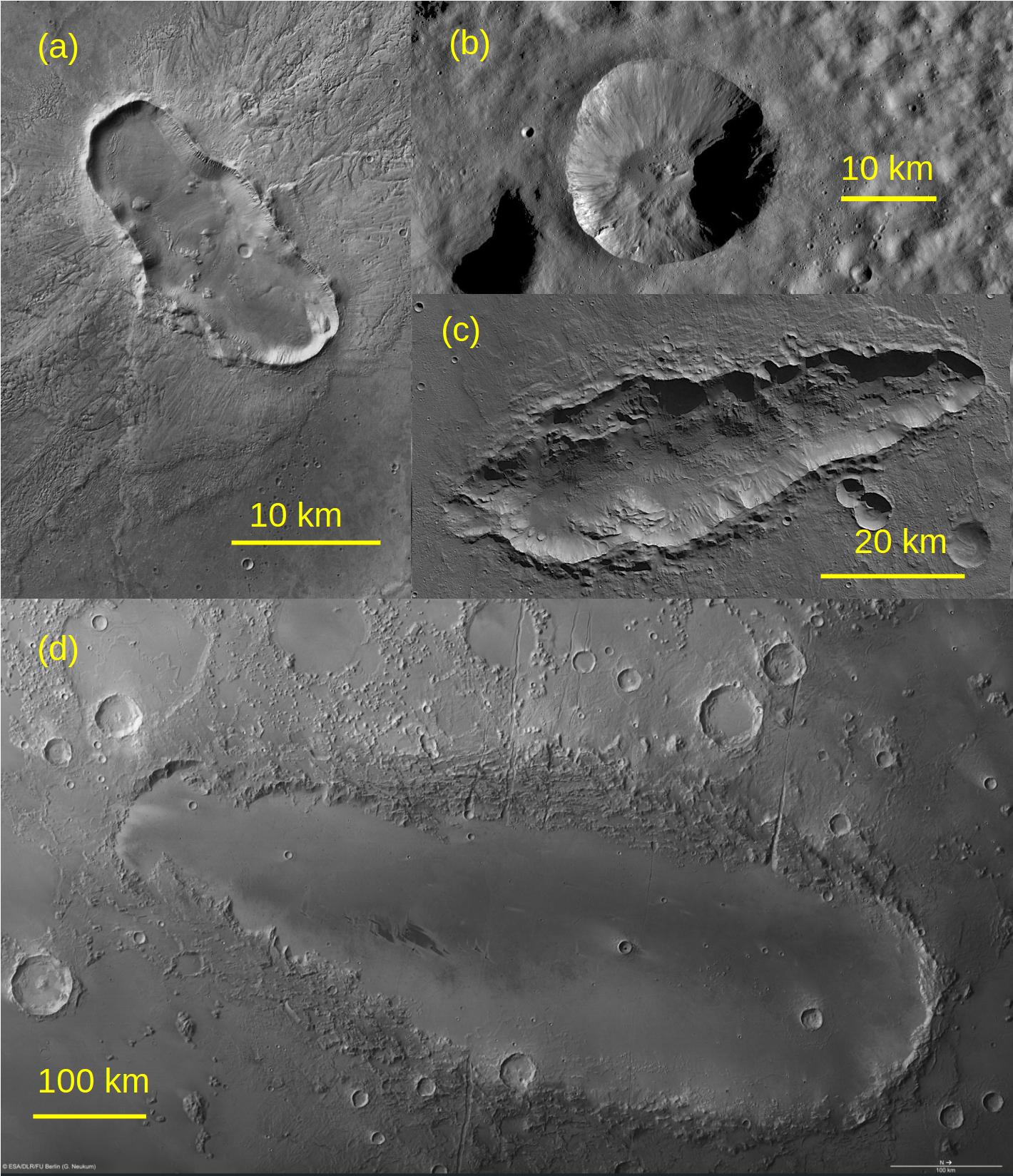}\\
	\end{center}
	\caption{\corr{(a) Butterfly crater in the region of Hesperia Planum, on Mars; (b) a recent 20-km-diameter bowl crater on Vesta; (c) elongated crater (78 km in length) in the south of the Huygens crater, on Mars; (d) Orcus Patera crater on Mars (with many circular craters in and around it). Image in panel (b): Courtesy NASA/JPL-Caltech. Images in panels (a), (c) and (d): Courtesy ESA/DLR/FU Berlin (G. Neukum).}}
	\label{fig:craters}
\end{figure}

The exact dynamics for distinct cratering processes differs, of course, with the amount of energy involved. Usually, very small energies imply partial penetration of the projectile, moderate energies can imply total penetration but also fragmentation of the projectile material \cite{Carvalho2}, and high-energy cratering involves melting and evaporation \cite{Ahrens, Pierazzo, Pierazzo2}. The diversity of scales and parameters makes of cratering an intricate problem, but the different systems share similarities if we consider only granular targets, which can then be explored to understand key features of crater formation. A dimensional analysis \cite{Holsapple} shows that the dimensionless volume of the crater is a function of the density ratio between the projectile $\rho_p$ and ground $\rho$ materials, $\rho_p/\rho$, and two pressure ratios. While one of the pressure ratios is important for the so-called strength regime, the other one is important in the so-called gravity regime and is defined as the projectile weight divided by its surface area and normalized by the dynamic pressure,

\begin{equation}
	\mathrm{Fr}^{-1} = \frac{D_p g}{V_p^2} \, ,
	\label{Eq:P}
\end{equation}

\noindent where $D_p$ is the projectile diameter, $V_p$ is the impact velocity, and $g$ is the modulus of gravity acceleration $\vec{g}$. The symbol Fr$^{-1}$ is used because Eq. (\ref{Eq:P}) corresponds to the inverse of a Froude number (gravitational to inertial effects). It has been shown that the ratio given by Eq. (\ref{Eq:P}) is the relevant dimensionless pressure when Fr$^{-1}$ $\lesssim$ 10$^{-2}$, in which fall great part of geophysical processes and collisions with cohesionless grains \cite{Holsapple}. Therefore, for the impact of projectiles onto a cohesionless granular soil, $\rho_p/\rho$ and Fr$^{-1}$ are considered as the two pertinent dimensionless groups. However, because $\rho_p/\rho$ is usually of order one, the results are frequently organized in terms of Fr$^{-1}$ only \cite{Holsapple}.

Besides the impact energy, other parameters can strongly influence the crater shape and projectile dynamics, such as confinement \cite{Seguin2}, ground composition \cite{Pacheco2}, projectile spin \cite{Carvalho}, microscopic friction \cite{Carvalho}, and bonding forces in aggregate-type projectiles \cite{Pacheco, Pacheco2, Carvalho2}. As a consequence, craters present very distinct shapes in nature \cite{Melosh, Pacheco2} and a comprehensive classification is still object of debate \cite{Arvidson, Barlow, Carvalho2}. Usually, small craters have a bowl shape, moderate to large craters can have a flat floor with a central peak and/or peak rings, and even larger craters have also external rings \cite{Melosh}, but the detailed dynamics for reaching each distinct shape is far from being completely understood. However, important hints were obtained, in especial for vertical impacts. For instance, different works on the vertical impact of solid projectiles showed that $\delta_y \sim D_{p}^{2/3}H^{1/3}$  \cite{Katsuragi2}, where $\delta_y$ is the projectile penetration depth and $H$ is the total vertical distance traveled by the projectile (Fig. \ref{fig:craters2}(a)). Although a lack of consensus in current penetration laws persists, some new clues were brought to light recently. By controlling the packing fraction $\phi$, Carvalho et al. \cite{Carvalho} showed that taking the packing fraction into account can, in fact, collapse existing laws. As an \textit{ad hoc} dependence, they proposed $\delta_y\phi^{n} \sim D_{p}^{2/3}H^{1/3}$, with $n$ = $9/2$ collapsing reasonably well data for different values of $\phi$. In terms of shape variability, Pacheco-V\'azquez and Ruiz-Su\'arez \cite{Pacheco2} showed that the vertical impact of aggregates generates flat craters with and without central peaks depending on the packing fraction of aggregates and impact velocity. More recently, Carvalho et al. \cite{Carvalho2} showed that, when spinning, aggregates impacting vertically a granular bed generate flat craters with central peak and peak rings if the angular velocity is relatively high and the bonding stresses within the agglomerate (sticking the grains together) are relatively small. In addition, they showed that if bonding stresses are moderate, spinning projectiles can generate asymmetric craters.

Another parameter that can greatly affect cratering is the angle of impact $\alpha$, that is, how inclined is the slanting trajectory of a projectile just before the impact takes place. Depending on $\alpha$, craters become more asymmetric, grains are ejected in preferential directions, and the projectile can rebound instead of being buried \cite{Gilbert, Gault, Pierazzo3}. Since impacts in nature (meteoroids and asteroids impacting planets and moons) are usually oblique \cite{Gilbert, Pierazzo3}, understanding the effects of $\alpha$ on cratering is of fundamental importance. Probably the first experiments inquiring into the effects of the impact angle of low-velocity projectiles on the formation of craters were those of Gilbert \cite{Gilbert}. His results showed that circular craters happens only for vertical impacts, while elongated craters take place when impacts are oblique. However, some decades later Gifford \cite{Gifford} showed that high-velocity impacts tend to generate circular craters due to the explosion caused during the impact, which would imply that craters found in nature should always be round. Gault and Wedekind \cite{Gault} showed later that, in fact, elongated and other asymmetric shapes can occur at high-velocity impacts if $\alpha$ $\lesssim$ 30$^{\circ}$. In that case, the crater morphology and the projectile fate depend on the impact velocity and the materials of target and projectile, with the asymmetry of ejecta deposits being noticeable, and the probability of ricochet becoming significant when $\alpha$ $\lesssim$ 15$^{\circ}$.  When $\alpha$ $\gtrsim$ 30$^{\circ}$, high-velocity impacts generate circular craters, which explains why most craters found on Mars and the Earth's moon are round (Gilbert \cite{Gilbert} and Shoemaker \cite{Shoemaker} had already shown that there is a probability of 75\% that the impacts of meteoroids and asteroids with planets and moons occur within 30$^{\circ}$ $\leq$ $\alpha$ $\leq$ 90$^{\circ}$ \cite{Pierazzo3}).

Recently, oblique impacts are receiving more attention due to their importance in nature \cite{Zheng, Wright, Bester, Takizawa, Ye, Ye2}. By carrying out experiments in a 2D (two dimensional) system with photoelastic particles, Bester et al. \cite{Bester} found that the inertial component of the drag force (due to collisional processes) decreases as $\alpha$ decreases, with force‐chain propagation in the horizontal plane becoming weaker. If the same weakening takes place in 3D systems, the projectile is then expected to migrate longer distances in the horizontal direction as the impact becomes more oblique (and to not rebound on the surface). Takizawa and Katsuragi \cite{Takizawa} investigated experimentally the simultaneous effects of the impact angle $\alpha$ and bed inclination on cratering, and proposed scaling laws based on dimensionless groups. Among other findings, they show that the crater volume follows scaling laws based on Fr$^{-1}$ and both the sine of the impact angle and the cosine of the target slope. For a sphere impacting onto a 3D granular system, Ye et al. \cite{Ye} found that both vertical and horizontal penetration depths scale linearly with impact velocity, and proposed a drag-force model that includes inertial, static, and viscous terms, the viscous contribution being significant for high impact velocities and deep penetration. Ye et al. \cite{Ye} also found that the projectile stopping time decreases with increasing $\alpha$ and is independent of impact velocity at fixed angle. Later, Ye et al. \cite{Ye2} investigated the ejecta splashing, corolla formation (curtain-like sheet of grains that splashes upward and outward), and shape of craters in oblique impacts, and found that crater shapes are well organized in a map in the Fr - $\alpha$ space. They also found that the bottom diameter of the corolla increases with $t^{1/2}$ and the ejecta height with $t$ in oblique impacts, where $t$ is time. In addition, they proposed a ballistic–avalanche model with air drag that successfully captures the evolution of the corolla and the final crater scaling.

Although previous studies advanced our understanding on different cratering mechanisms and resulting craters, some questions remain open. For example, what are the relative roles of rotation and impacting angle in the shape of resulting craters and trajectory of the projectile? Is rebound mostly due to initial spinning of the projectile or to the impacting angle? Are there preferential directions of the initial spin for penetration? In this paper, we inquire into these questions by carrying out numerical computations using 3D DEM (three-dimensional discrete element method). In our simulations, projectiles with an initial spin impacted a cohesionless granular bed, and we varied the magnitude and direction of the projectile spin, the impact velocity, the bed packing fraction, and the incident angle. We show that the projectile can rebound (ricochet) for small angles, or be completely or partially buried for larger angles, and that when buried it can sometimes migrate large horizontal distances depending on the incident angle. We also show that increasing the packing fraction strengthens rebounds, and that the initial spin induces rebound, burying, or transverse deviations depending on its direction and orientation. In particular, when the spin is in the transverse direction, a lifting force appears with inverted sense with respect to the Magnus force found in fluids. The crater morphology also changes with the varying parameters, acquiring circular, elliptical, drop-like, tadpole-like, and transitional shapes, and we find that the shapes correlate well with the projectile trajectory. Finally, we propose diagrams organizing and classifying the impact dynamics, from which we can observe the relative roles of the impacting angle, initial spin, and packing fraction on the cratering process. Our results shed new light on crater morphology and fate of the impacting material.

\section{MODEL AND NUMERICAL SETUP}
\label{sec:mumerical}

\begin{figure}[h!]
	\begin{center}
		\includegraphics[width=0.99\linewidth]{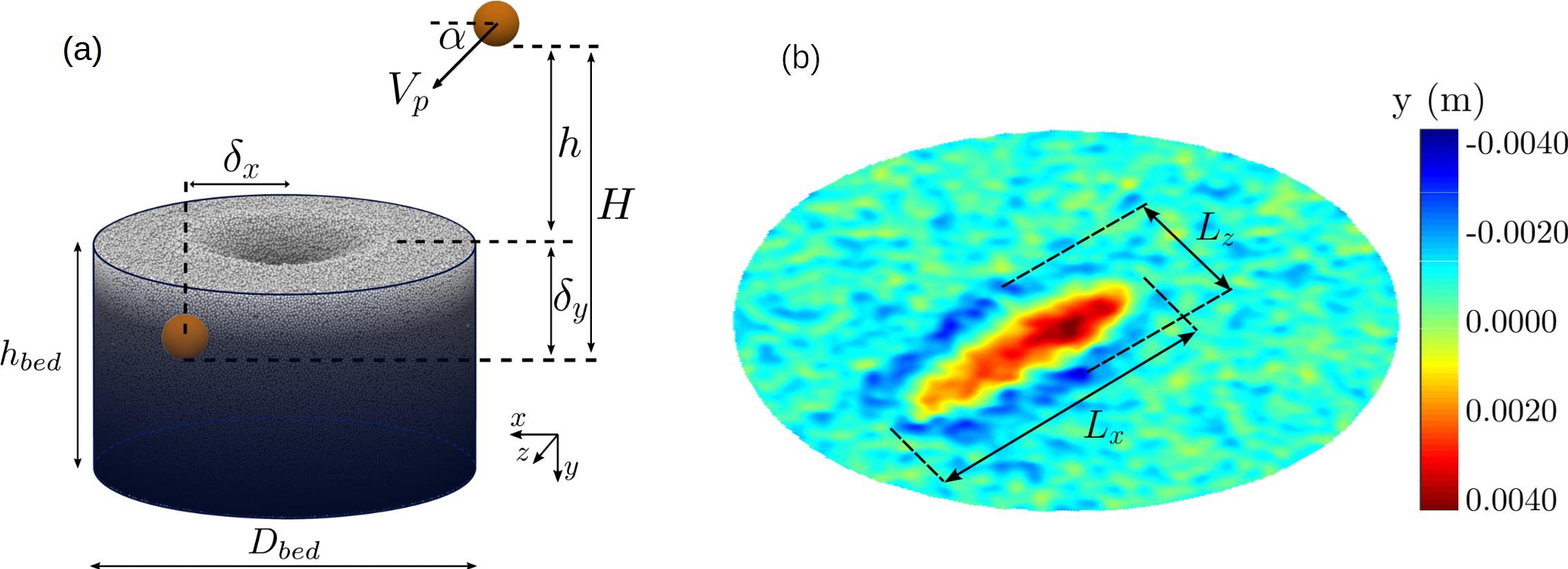}\\
	\end{center}
	\caption{\corr{(a) Layout of the numerical setup (the $y$ coordinate points downwards, and, although shown on the bottom, the origin of the coordinate system is on the bed surface centered horizontally in the domain). (b) Numerical result showing the topography (elevation) of a crater formed by an oblique impact. In this figure, $\phi$ $=$ 0.559, $\alpha$ $=$ 5$^\circ$, $V_p$ $=$ 6.5 m/s, and the colorbar shows the elevation from the undisturbed surface (pointing downwards). The longitudinal $L_x$ and transverse $L_z$ dimensions of the crater are shown in the figure.}}
	\label{fig:craters2}
\end{figure}

We carried out 3D DEM computations \cite{Cundall} of a projectile impacting onto a bed of spheres at different angles $\alpha$ (Fig. \ref{fig:craters2}(a)), linear velocities $V_p$, and initial spins (angular velocities) $\vec{\omega}$. For that, we made use of the open-source code LIGGGHTS \cite{Kloss2, Berger}, which solves the linear (Eq. (\ref{Fp})) and angular (Eq. (\ref{Tp})) momentum equations for each individual particle,

\begin{equation}
	m\frac{d\vec{u}}{dt}= \vec{F}_{c} + m\vec{g} \,\, ,
	\label{Fp}
\end{equation}

\begin{equation}
	I\frac{d\vec{\omega}}{dt}=\vec{T}_{c} \,\, ,
	\label{Tp}
\end{equation}

\noindent where $m$ is the mass of the considered particle, $\vec{u}$ is its velocity, $I$ is its moment of inertia, $\vec{\omega}$ is its angular velocity, $\vec{F}_{c}$ is the resultant of contact forces acting on the considered particle, and $\vec{T}_{c}$ is the resultant of contact torques. The resultant of contact forces $\vec{F}_{c}$ and torques $\vec{T}_{c}$ are computed by Eqs. (\ref{Fc}) and (\ref{Tc}), respectively,

\begin{equation}
	\vec{F}_{c} = \sum_{i\neq j}^{N_c} \left(\vec{F}_{c,ij} \right) + \sum_{i}^{N_w} \left( \vec{F}_{c,iw} \right) \,\,,
	\label{Fc}
\end{equation}

\begin{equation}
	\vec{T}_{c} = \sum_{i\neq j}^{N_c} \vec{T}_{c,ij} + \sum_{i}^{N_w} \vec{T}_{c,iw} \,\,,
	\label{Tc}
\end{equation}

\noindent \corr{where $\vec{F}_{c,ij}$ and $\vec{F}_{c,iw}$} are the contact forces between particles $i$ and $j$ and between particle $i$ and the wall, respectively, $\vec{T}_{c,ij}$ and $\vec{T}_{c,iw}$ are torques due to the tangential component of the contact forces between particles $i$ and $j$ and between particle $i$ and the wall (both considering rolling resistance), respectively, $N_c$ - 1 is the number of particles in contact with particle $i$, and $N_w$ is the number of particles in contact with the wall. The contact forces and torques are computed using the elastic Hertz-Mindlin contact model \cite{direnzo}, and the rolling resistance is taken into account by using a coefficient of rolling friction \cite{Derakhshani} (see Ref. \cite{Carvalho} for a detailed description of the DEM model used).

The numerical domain consisted of a granular bed in a cylindrical container and a projectile with diameter $D_p$ $=$ 0.025 m and density $\rho_p$ $=$ 7865 kg/m$^3$. The granular bed had a diameter $D_{bed}$ = 400 mm and height $h_{bed}$, consisting of $N$ $\approx$ 10$^6$ spheres with diameter within 2.1 mm $\leq$ $d$ $\leq$ 2.9 mm (following a Gaussian distribution) and density $\rho$ = 2600 kg/m$^3$. Prior to each simulation, 1.2 $\times$ 10$^6$ spheres randomly arranged in space were let to fall freely and settle in the container until a low level of kinetic energy was attained. In doing that, different values of the average packing fraction $\phi$ could be obtained by varying the initial value of the coefficient of grain-grain friction $\mu_{gg}$, and then changing it back to the correct value. For assuring a horizontal surface, grains at a height above $h_{bed}$ were deleted, which corresponded to approximately 10$^4$ grains, keeping then $N$ $\approx$ 10$^6$. With this procedure, $h_{bed}$ = 118-129 mm, depending on the average packing fraction.  The distribution of diameters used in the simulations are shown in Tab. \ref{tab_diameters}, and the properties used for the grains and projectile (taken from the literature) are listed in Tabs. \ref{tabmaterials} and \ref{tabcoefficients}. \corr{We note that the value of the Young's modulus $E$ adopted in our simulations for the steel projectile is one order of magnitude smaller than the real one ($E$ $=$ 1.96 $\times$ 10$^{11}$ Pa), since it allows the use of higher time steps while keeping a reasonable accuracy \cite{Lommen}. For the steel walls, we used the real value.}

\begin{table}[!h]
	\centering
	\caption{Distribution of diameters for the settling grains: number of grains $N_d$ of each diameter $d$, for the different packing fractions $\phi$ simulated.}
	\label{tab_diameters}
	\begin{tabular}{|c|c|c|c|c|c|}
		\hline
		$d$ (mm) & 2.1 & 2.3 & 2.5 & 2.7 & 2.9 \\
		\hline
		$N_d$ ($\phi$ = 0.559)  & 25026 & 149566 & 750128 & 149092 & 24890\\
		\hline
		$N_d$ ($\phi$ = 0.590)  & 25061 & 150035 & 753921 & 150285 & 25120\\
		\hline
		$N_d$ ($\phi$ = 0.638)  & 26062 & 155621 & 782328 & 156025 & 26111\\
		\hline
	\end{tabular}
\end{table}

\begin{table}[!h]
	\centering
	\caption{Properties of materials used in the simulations: $E$ is Young's modulus, $\nu$ is the Poisson ratio, and $\rho$ is the material density. The last column corresponds to the diameter of the considered object.}
	\label{tabmaterials}
	\begin{tabular}{l|c|c|c|c|c}
		\hline
		& \textbf{Material} & \textbf{$E$ (Pa)} & \textbf{$\nu$} & \textbf{$\rho$ (kg/m$^{3}$)}& \textbf{Diameters (mm)}\\
		\hline
		Projectile & Steel\footnotesize{$^{(1)}$} & $0.2 \times 10^{11}$  & 0.3 & 7865 & 25\\
		Grains & Sand\footnotesize{$^{(1)-(2)}$} & $0.1 \times 10^{9}$ & 0.3 & 2600 & 2.1 $\leq$ $d$ $\leq$ 2.9 \\
		Walls & Steel \footnotesize{$^{(1)}$} & $0.2 \times 10^{12}$ & 0.3 & 7865 & 400\\    
		\hline
		\multicolumn{3}{l}{\footnotesize{$^{(1)}$ Ucgul et al. \cite{Ucgul1, Ucgul2, Ucgul3}}} \\
		\multicolumn{3}{l}{\footnotesize{$^{(2)}$ Derakhshani et al. \cite{Derakhshani}}}\\		
	\end{tabular}
\end{table}

\begin{table}[!h]
	\centering
	\caption{Coefficients used in the numerical simulations.}
	\label{tabcoefficients}
	\begin{tabular}{l|c|c}
		\hline
		\textbf{Coefficient}  & \textbf{Symbol} & \textbf{Value} \\
		\hline		  
		Restitution coefficient (grain-grain)\footnotesize{$^{(1)}$} & $\epsilon_{gg}$ & 0.6 \\
		Restitution coefficient (grain-projectile)\footnotesize{$^{(1)}$} & $\epsilon_{gp}$ & 0.6 \\
		Restitution coefficient (grain-wall)\footnotesize{$^{(1)}$} & $\epsilon_{gw}$ & 0.6 \\
		Fiction coefficient (grain-grain)\footnotesize{$^{(1)-(2)}$} & $\mu_{gg}$ & 0.52 \\
		Friction coefficient (grain-projectile)\footnotesize{$^{(1)}$} & $\mu_{gp}$ & 0.5 \\
		Friction coefficient (grain-walls)\footnotesize{$^{(1)}$} & $\mu_{gw}$ & 0.5 \\		
		Coefficient of rolling friction (grain-grain)\footnotesize{$^{(2)}$} & $\mu_{r,gg}$ & 0.3\\
		Coefficient of rolling friction (grain-projectile)\footnotesize{$^{(1)}$} & $\mu_{r,gp}$ & 0.05\\
		Coefficient of rolling friction (grain-wall)\footnotesize{$^{(1)}$} & $\mu_{r,gw}$ & 0.05\\
		\hline
		\multicolumn{3}{l}{\footnotesize{$^{(1)}$ Ucgul et al. \cite{Ucgul1, Ucgul2, Ucgul3}}} \\
		\multicolumn{3}{l}{\footnotesize{$^{(2)}$ Derakhshani et al. \cite{Derakhshani}}}
	\end{tabular}
\end{table}

For the simulations, we imposed a collision velocity to the projectile $\vec{V_p} = V_{p} cos(\alpha) \vec{i} + V_{p} sin(\alpha) \vec{j}$, where $V_p = |\vec{V_p}|$, with $V_p$ $=$ $\sqrt{2gh}$ corresponding to a free-fall height $h$. In our simulations, Froude numbers were within 0.04 $\leq$ Fr $\leq$ 400 (2.5 $\times$ 10$^{-3}$ $\leq$ Fr$^{-1}$ $\leq$ 2.5 $\times$ 10), and we used a time step $\Delta t$ $=$ 2 $\times$ 10$^{-6}$ s, which corresponds to less than 10 \% of the Rayleigh time \cite{Derakhshani}. Figure \ref{fig:craters2}(a) shows a layout of the numerical setup, and animations showing the different types of impacts are available in the Supplemental Material \cite{Supplemental}.  The numerical setup of our simulations, output files, and scripts for post-processing the outputs are available in an open repository \cite{Supplemental2}.

\section{\label{sec:Results} RESULTS AND DISCUSSION}

\subsection{\label{general_observations} General observations}

\begin{figure}[ht]
	\begin{center}
		\includegraphics[width=\linewidth]{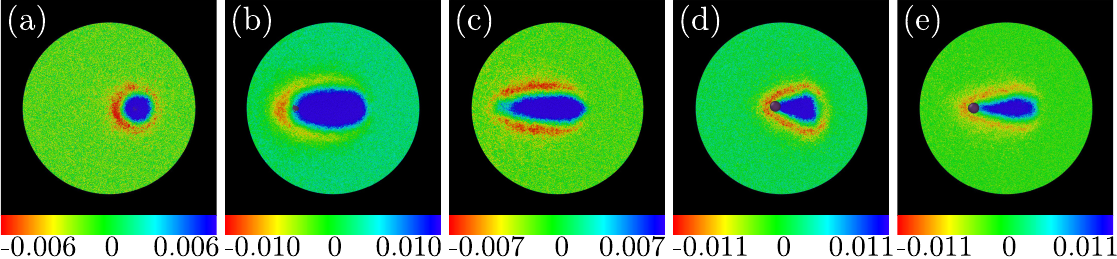}\\
	\end{center}
	\caption{Top view of final positions of grains, showing the different shapes of craters found by varying the controlled parameters. (a) Circle, obtained in this figure for $\alpha$ $=$ 75$^\circ$, $V_p$ $=$ 4 m/s, and $\phi$ $=$ 0.559; (b) ellipse, obtained in this figure for $\alpha$ $=$ 30$^\circ$, $V_p$ $=$ 9 m/s, and $\phi$ $=$ 559; (c) tadpole-like crater, obtained in this figure for $\alpha$ $=$ 15$^\circ$, $V_p$ $=$ 7 m/s, and $\phi$ $=$ 0.559; (d) goutte-like crater, obtained in this figure for $\alpha$ $=$ 60$^\circ$, $V_p$ $=$ 5 m/s, and $\phi$ $=$ 0.638. (e) transitional shape, obtained in this figure for $\alpha$ $=$ 30$^\circ$, $V_p$ $=$ 4 m/s, and $\phi$ $=$ 0.590. In all panels, $|\vec{\omega}|$ $=$ 0 rad/s. The colorbar on the bottom of each panel shows the elevation of each grain from the undisturbed surface (coordinate pointing downwards), in meters. }
	\label{fig:morphology}
\end{figure}

\begin{figure}[ht]
	\begin{center}
		\includegraphics[width=\linewidth]{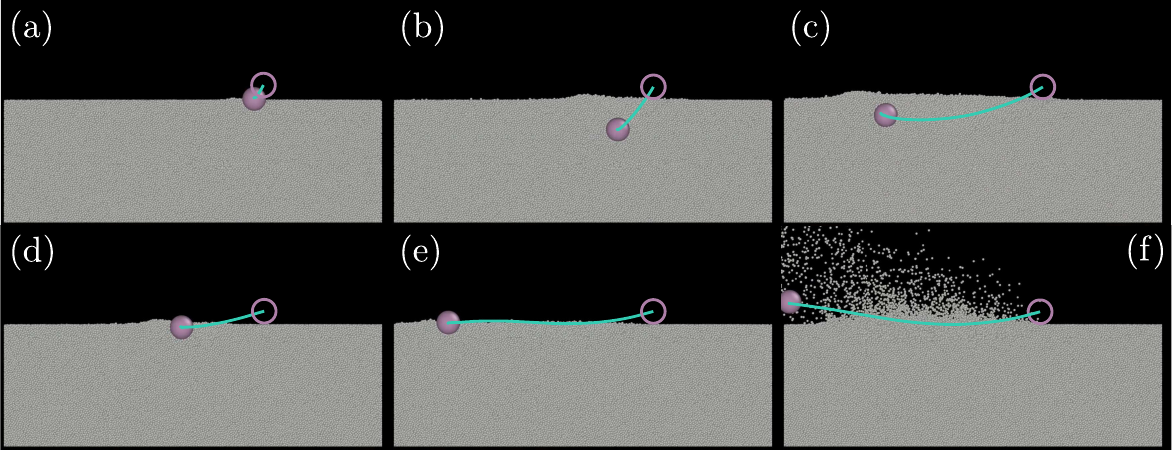}\\
	\end{center}
	\caption{Different fates of the projectile observed in the DEM computations: (a) partial burying, obtained in this figure for $\alpha$ $=$ 60$^\circ$, $V_p$ $=$ 0.5 m/s, and $\phi$ $=$ 0.559; (b) penetration, obtained in this figure for $\alpha$ $=$ 60$^\circ$, $V_p$ $=$ 4 m/s, and $\phi$ $=$ 0.559; (c) subsurface glide, obtained in this figure for $\alpha$ $=$ 30$^\circ$, $V_p$ $=$ 9 m/s, and $\phi$ $=$ 0.559; (d) subsurface rise, obtained in this figure for $\alpha$ $=$ 15$^\circ$, $V_p$ $=$ 1.5 m/s, and $\phi$ $=$ 0.559; (e) retained ricochet, obtained in this figure for $\alpha$ $=$ 15$^\circ$, $V_p$ $=$ 3 m/s, and $\phi$ $=$ 0.559; (f) ricochet, obtained in this figure for $\alpha$ $=$ 15$^\circ$, $V_p$ $=$ 8 m/s, and $\phi$ $=$ 0.559. In all panels, $|\vec{\omega}|$ $=$ 0 rad/s. The impact point is shown as an empty circle, the final position as a solid circle, and the trajectory as a solid line.}
	\label{fig:fate_projectile}
\end{figure}

From our numerical computations, we have access to the positions of the projectile and all grains along the cratering process, so that we can analyze both the morphology of craters and the trajectory of the projectile. Some examples of the final shapes of craters and behavior of the projectile are shown in Figs. \ref{fig:morphology} and \ref{fig:fate_projectile}, respectively. A movie showing the cratering process and the trajectory of the projectile is available in the Supplemental Material \cite{Supplemental}. 
  
Figure \ref{fig:morphology} presents top view images of the final positions of grains, showing the different shapes of craters found by varying $\alpha$, $V_p$, and $\phi$ (the values used in each panel are listed in the figure caption). We observe that craters can acquire one of the following shapes:

\begin{itemize}
	\item circle (Fig. \ref{fig:morphology}(a));
	\item ellipse (Fig. \ref{fig:morphology}(b));
	\item tadpole-like shape (Fig. \ref{fig:morphology}(c));
	\item goutte-like shape (Fig. \ref{fig:morphology}(d));
	\item transitional shape (shape between the tadpole- and goutte-like shapes, Fig. \ref{fig:morphology}(e)).
\end{itemize}

\noindent The first two shapes are reported in the literature to appear, respectively, for large and relatively low values of $\alpha$, while the last three shapes can appear for relatively large or low incident angles, depending on the values of $V_p$, $\phi$ and $\vec{\omega}$. We remark the similarities between the tadpole-like \corr{(Fig. \ref{fig:morphology}(c))} and transitional \corr{(Fig. \ref{fig:morphology}(e))} craters and those in Figs. \ref{fig:craters}(c) and \ref{fig:craters}(d) found on Mars, the conditions for such shapes being described next in Subsection \ref{morphology}. \corr{Of course, there is also a strong similarity between the circle crater of Fig. \ref{fig:morphology}(a) and the bowl crater showed in Fig. \ref{fig:craters}(b), while a slighter similarity is found between the elliptical crater of Fig. \ref{fig:morphology}(b) and the butterfly crater showed in Fig. \ref{fig:craters}(a).}

The projectile behavior is the responsible for the excavation of the crater, and its trajectory depends on the values of $\alpha$, $V_p$, $\phi$, and $\vec{\omega}$. Figure \ref{fig:fate_projectile} shows the initial position, trajectory, and final position (fate) of the projectile, for the different fates observed in the numerical simulations. The projectile fate can be classified in the following categories:

\begin{itemize}
	\item partial burying (partial penetration of the projectile, Fig. \ref{fig:fate_projectile}(a));
	\item penetration (Fig. \ref{fig:fate_projectile}(b));
	\item subsurface glide (the projectile penetrates the bed and afterward migrates a considerable distance in the horizontal direction, Fig. \ref{fig:fate_projectile}(c));
	\item subsurface rise (similar to the subsurface glide, but the projectile rises while migrating horizontally, Fig. \ref{fig:fate_projectile}(d));
	\item retained ricochet (the projectile initiates a rebound at the surface, but remains retained by grains bulldozed at its front, Fig. \ref{fig:fate_projectile}(e));
	\item ricochet (rebound, Fig. \ref{fig:fate_projectile}(f)).
\end{itemize}

\noindent \corr{The classification of craters is based on the second-moment equivalent ellipse and thickness-to-length refinement, and that of fates makes use of an outcome-based decision tree. The procedures adopted for classifying the crater morphologies and projectile fates are described in detail in the Supplemental Material \cite{Supplemental}.}

There is a correlation between the projectile fate and crater shapes, as we show later in Subsection \ref{correlation}. But first, we inquire into the morphology of craters and fate of the projectile in Subsections \ref{morphology} and \ref{fate}.

\subsection{\label{morphology} Crater morphology}

\begin{figure}[ht]
	\begin{center}
		\includegraphics[width=\linewidth]{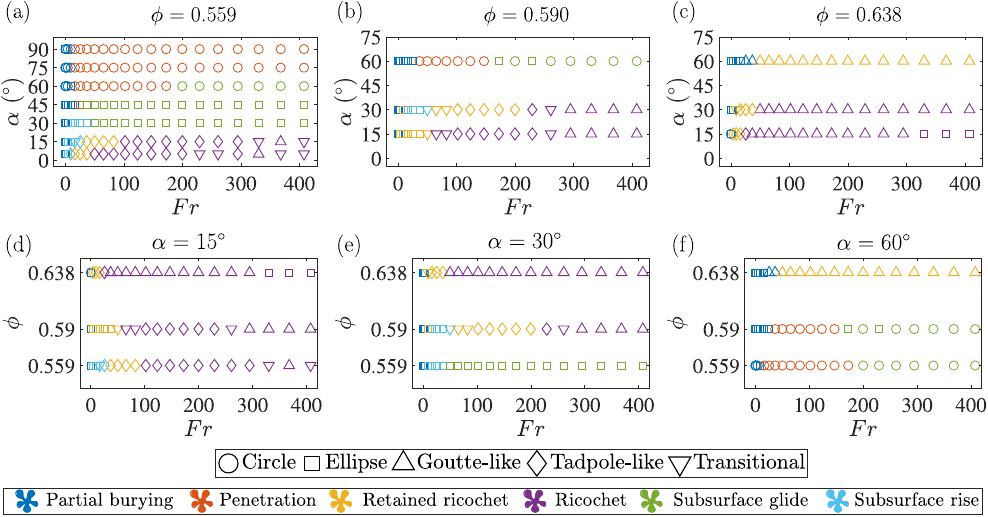}\\
	\end{center}
	\caption{Classification maps of crater shapes for non-spinning projectiles. (a)-(c) Maps in the $\alpha$ - Fr space for different packing fractions: $\phi$ $=$ 0.559, 0.590 and 0.638, respectively. (d)-(f) Maps in the $\phi$ - Fr space for different impacting angles: $\alpha$ $=$ 15$^{\circ}$, 30$^{\circ}$ and 60$^{\circ}$, respectively. The symbols are listed in the key, and the colors \corr{(listed also in the key)} correspond to the projectile fates presented in Subsection \ref{fate}.}
	\label{fig:morphology2}
\end{figure}

\begin{figure}[ht]
	\begin{center}
		\includegraphics[width=\linewidth]{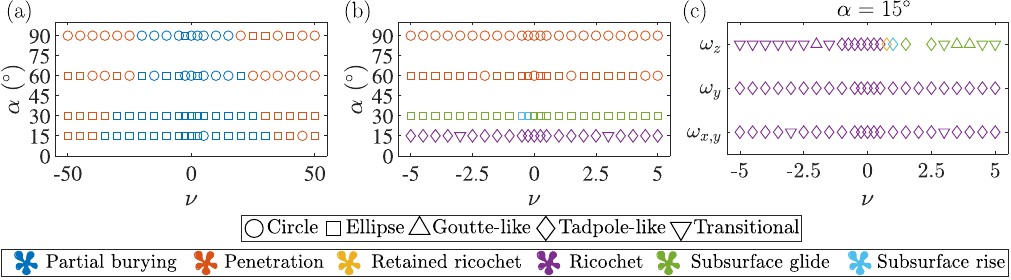}\\
	\end{center}
	\caption{Classification maps of crater shapes for spinning projectiles. (a) and (b) Maps in the $\alpha$ - $\nu$ space for $\phi$ $=$ 0.559 and different impacting velocities: $V_p$ $=$ 0.5 and 5 m/s, respectively, corresponding to Fr $=$ 1 and 102. (c) Map in the $\omega$ direction - $\nu$ space, for $\alpha$ $=$ 15$^{\circ}$ and $V_p$ $=$ 5 m/s (Fr $=$ 102). The symbols are listed in the key, and the colors \corr{(listed also in the key)} correspond to the projectile fates presented in Subsection \ref{fate}.}
	\label{fig:morphology3}
\end{figure}

We begin evaluating how crater shapes change with $\alpha$, $V_p$, $\phi$, and $\vec{\omega}$. For that, we carried out extensive simulations in which we varied one of those parameters while the others remained fixed, and the results are summarized in Figs. \ref{fig:morphology2} and \ref{fig:morphology3}. Figures \ref{fig:morphology2}(a)-(c) show maps of crater shapes in the $\alpha$ - Fr space for different packing fractions ($\phi$ $=$0.559, 0.590 and 0.638, respectively), and Figs. \ref{fig:morphology2}(d)-(f) show the maps in the $\phi$ - Fr space for different impacting angles ($\alpha$ $=$ 15$^{\circ}$, 30$^{\circ}$ and 60$^{\circ}$, respectively), all figures corresponding to non-spinning projectiles ($\vec{\omega}$ $=$ 0). Concerning spinning projectiles, Figs. \ref{fig:morphology3}(a) and \ref{fig:morphology3}(b) show maps of crater shapes in the $\alpha$ - $\nu$ space for $\phi$ $=$ 0.559 and different impacting velocities ($V_p$ $=$ 0.5 and 5 m/s, respectively, corresponding to Fr $=$ 1 and 102), and Fig. \ref{fig:morphology3}(c) shows a classification map in the $\omega$ direction - $\nu$ space for $\alpha$ $=$ 15$^{\circ}$ and $V_p$ $=$ 5 m/s. In these Figures, $\nu$ is the spin normalized by the grain diameter $d$ and impact velocity $V_p$, corresponding, thus, to the square root of the ratio between the rotational and linear kinetic energies of the projectile \cite{Carvalho3},

\begin{equation}
	\nu= sign(\vec{\omega}) \frac{|\vec{\omega}| d}{V_p} \,\,.
	\label{eq:nu}
\end{equation}

\noindent The colors of symbols in Figs. \ref{fig:morphology2} and \ref{fig:morphology3} describe the projectile fate as shown in Fig. \ref{fig:fate2} (same colors). The components of $\vec{\omega}$ analyzed are those aligned in the transverse direction, $\omega_z$, in the vertical direction, $\omega_y$, and in the impact direction, $\omega_{x,y}$, where,

\begin{equation}
	\vec{\omega}_{x,y} = \omega_{x,y} \cos(\alpha) \vec{i} + \omega_{x,y} \sin(\alpha) \vec{j} \,\,,
\end{equation}

\noindent in which $|\vec{\omega}_{x,y}|$ $=$ $\omega_{x,y}$, and the $sign(\vec{\omega})$ in Eq. (\ref{eq:nu}) corresponds to the sign of the \textit{j} component of $\vec{\omega}_{x,y}$ in this specific case.

For a non-spinning projectile, Fig. \ref{fig:morphology2}(a) shows that when $\phi$ $=$ 0.559 (typical of random loose beds) circle craters appear for $\alpha$ $\gtrsim$ 50$^{\circ}$, irrespective of Fr, while elliptical craters appear for  20$^{\circ}$ $\lesssim$ $\alpha$ $\lesssim$ 50$^{\circ}$, also irrespective of Fr. For $\alpha$ $\lesssim$ 20$^{\circ}$, elliptical craters appear for small Fr (with the band increasing with $\alpha$), tadpole-like craters appear for 10 $\lesssim$ Fr $\lesssim$ 350, and the transitional and goutte-like craters appear for Fr $\gtrsim$ 200. These results follow the trend reported \corr{in the literature, even} for high-speed impacts \cite{Gault}, although the values of threshold angles can be different (in case of high speeds due to differences in energy, and in the other cases due to the values of $\phi$, which are not even reported in many of the experiments). \corr{For instance, our diagram for $\phi$ $=$ 0.559 shows reasonable agreement with Fig. 3 of Ye et al. \cite{Ye2}, noting that in their work they used a different classification (triangle and rectangular craters instead of goutte-like, tadpole-like and transitional craters), and that they had an uncertainty of 4\% in the packing fraction and a different grain polidispersity ($\pm$ 40\%) than ours. With that, they found different values for the limits between crater shapes, but their distributions in the $\alpha$ - $\nu$ space resemble ours. Interestingly, Ye et al. \cite{Ye2} found that the transition lines between different crater shapes in the $\alpha$ - $\nu$ space follow approximately exponential curves, but we cannot make the same assertion from our data.}

Different from previous works, we can control the exact values of the mean packing fraction, and, thus, investigate its effects on impact cratering. For higher packing fractions, simulations using $\phi$ $=$ 0.590 show that circle craters still occur for $\alpha$ $=$ 60$^{\circ}$, but now competing with elliptical craters, and when $\phi$ $=$ 0.638 circle craters are not observed for $\alpha$ $=$ 60$^{\circ}$, elliptical and goutte-like craters appearing instead. For $\phi$ $=$ 0.590 and 0.638 and $\alpha$ $<$ 60$^{\circ}$, elliptical, tadpole-like, goutte-like, and transitional craters appear, the goutte-like shape dominating when $\phi$ $=$ 0.638 (typical of random close packing). \corr{For better visualizing the distribution of crater shapes as a function of the packing fraction, Figs. \ref{fig:morphology2}(d)-(f) show part of the results in the $\phi$ - Fr space for three different impacting angles: $\alpha$ $=$ 15$^{\circ}$, 30$^{\circ}$ and 60$^{\circ}$, respectively. We can clearly observe the lack of circle craters when $\alpha$ $<$ 60$^{\circ}$ (just observed for the smaller Fr number and higher packing fraction, for which the impact velocity is low and partial burying occurs), and the existence of circle and elliptical craters when $\alpha$ $>$ 60$^{\circ}$ and $\phi$ $=$ 0.559 and 0.590}.

We further extended our investigation by imposing an initial spin to the projectile, and varying its intensity and direction. For example, for a spinning projectile with angular velocity aligned with the impact direction ($\omega_{x,y}$), Fig. \ref{fig:morphology3}(a) shows that circular and elliptical craters can coexist for $\alpha$ $\gtrsim$ 60$^{\circ}$ when the impact velocity is relatively low ($V_p$ $=$ 0.5 m/s). For  $\alpha$ $\lesssim$ 60$^{\circ}$, elliptical craters dominate. For all angles, partial burying is observed for relatively small values of $\nu$ (relatively small spins), while penetration is observed for higher values of $\nu$ (the values of the threshold decrese with increasing the angle).  When the impact velocity $V_p$ increases by one order of magnitude ($V_p$ $=$ 5 m/s), Fig. \ref{fig:morphology3}(b) shows that there are only circle craters for $\alpha$ $=$ 90$^{\circ}$ (normal impacts), a majority of elliptical craters (coexisting with circle craters) for $\alpha$ $=$ 60$^{\circ}$, only elliptical craters when $\alpha$ $=$ 30$^{\circ}$, and tadpole-like and transitional craters when $\alpha$ $=$ 15$^{\circ}$. Under this impact velocity, the projectile fate is almost independent of the initial spin, with penetration occurring for $\alpha$ $\gtrsim$ 60$^{\circ}$, subsurface glide and rise occurring for $\alpha$ $=$ 30$^{\circ}$, and ricochet for $\alpha$ $=$ 15$^{\circ}$. For the specific case when $\alpha$ $=$ 15$^{\circ}$ and $V_p$ $=$ 5 m/s, Fig. \ref{fig:morphology3}(c) shows that the transverse component of the spin, $\omega_z$, gives rise to tadpole-like craters for small values of the spin, and to transitional  craters for higher spin values. For negative values of $\omega_z$, we observe ricochet (since the tangential velocity of the projectile at the contact points rearwards, adding to the projectile translation), and for positive values retained ricochet, subsurface rise, and subsurface glide (by augmenting $\nu$). For the vertical and longitudinal (aligned with the impacting velocity) components of the spin ($\omega_y$ and $\omega_{x,y}$, respectively), craters are mostly tadpole like, with ricochet always taking place.

\begin{figure}[ht]
	\begin{center}
		\includegraphics[width=0.99\linewidth]{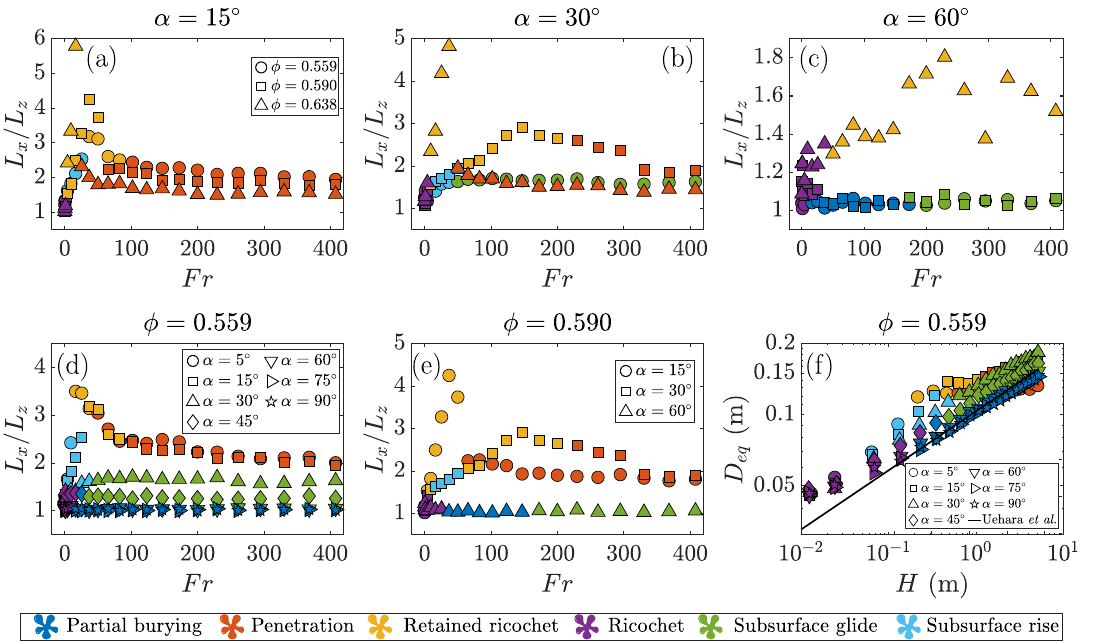}\\
	\end{center}
	\caption{Aspect ratio ($L_x/L_z$) and equivalent diameter $D_{eq}$ of craters. (a)-(c) $L_x/L_z$ as a function of Fr parameterized by the packing fraction $\phi$, for $\alpha$ $=$ 15$^{\circ}$, 30$^{\circ}$ and 60$^{\circ}$, respectively. (d) and (e) $L_x/L_z$ as a function of Fr parameterized by $\alpha$, for $\phi$ $=$ 0.559 and 0.590, respectively. (f)  $D_{eq}$ as a function of the total vertical distance $H$, for $\phi$ $=$ 0.559. The symbols are listed in the keys, and the colors \corr{(listed also in the key)} correspond to the projectile fates presented in Subsection \ref{fate}.}
	\label{fig:morphology4}
\end{figure}

In order to better quantify the morphology, we computed the aspect ratio of shapes as the ratio between the longitudinal distance $L_x$ and the larger transverse distance $L_z$ of the craters. In addition, we computed an equivalent diameter $D_{eq}$ given by four times the planar area (horizontal plane) divided by the perimeter of craters, and both results are shown in Fig. \ref{fig:morphology4}. Figures \ref{fig:morphology4}(a)-(c) show $L_x/L_z$ as a function of Fr, parameterized by the packing fraction $\phi$, for $\alpha$ $=$ 15$^{\circ}$, 30$^{\circ}$ and 60$^{\circ}$, respectively. For the smallest impacting angle ($\alpha$ $=$ 15$^{\circ}$), Fig. \ref{fig:morphology4}(a) shows a non-monotonic behavior, with the elongation (asphericity) initially increasing with Fr until reaching a maximum, from which point it decreases and tends to a plateau. The initial increase of elongation occurs until Fr $\lesssim$ 50, with subsurface rise, ricochet, and mainly retained ricochet taking place, while the decrease occurs with retained ricochet and penetration happening at larger values of Fr. The maximum value of elongation occurs under retained ricochet, which is coherent with the fact that the projectile undergoing retained ricochet moves a considerable distance on the surface. Interestingly, the maximum elongation increases with $\phi$, and this is probably due to a less deep penetration of the projectile during the retained ricochet process when $\phi$ is higher. By increasing $\alpha$ (Figs. \ref{fig:morphology4}(b) and \ref{fig:morphology4}(c)), we observe an initial increase only for the higher values of $\phi$ ($\phi$ $=$ 0.590 and 0.638 when $\alpha$ $=$ 30$^{\circ}$, and $\phi$ $=$ 0.638 when $\alpha$ $=$ 60$^{\circ}$), with the region of increasing elongation with Fr occurring under subsurface rise, ricochet, and mainly retained ricochet. The maximum values are always obtained for retained ricochet, at Fr $\approx$ 50 when $\alpha$ $=$ 30$^{\circ}$ and Fr $\approx$ 200 when $\alpha$ $=$ 60$^{\circ}$. In the plateau region, we observe subsurface glide, partial burying, and penetration. We also observe that craters tend to be less elongated as $\alpha$ increases, mainly for the smaller packing fractions $\phi$ tested. This is shown in Figs. \ref{fig:morphology4}(b) and \ref{fig:morphology4}(c), from which we notice that the maximum values of $L_x/L_z$ decrease and the peak tends to disappear as $\alpha$ increases, with lower elongations for $\phi$ $=$ 0.559 and 0.590. These observations are corroborated by Figs. \ref{fig:morphology4}(d) and \ref{fig:morphology4}(e), which show $L_x/L_z$ as a function of Fr parameterized by $\alpha$, for $\phi$ $=$ 0.559 and 0.590, respectively. From these figures, we observe larger elongations when $\alpha$ is small and Fr is within a moderate range. Finally, Fig. \ref{fig:morphology4}(f) shows the equivalent diameter $D_{eq}$ as a function of the total vertical distance $H$ $=$ $h$ $+$ $\delta_y$ traveled by the projectile. We observe that, when $\alpha$ $\geq$ 45$^{\circ}$, $D_{eq}$ varies with $H$ following approximately the correlation proposed by Uehara et al. \cite{Uehara, Uehara2} for the diameter $D_c$ of circular craters,

\begin{equation}
	D_{c} = 0.90\bigg(\frac{\rho_{p}}{\rho \mu_{rep}^2}\bigg)^{1/4}D_{p}^{3/4}H^{1/4} \,,
	\label{Dc_uehara}
\end{equation}

\noindent even if the present craters are elongated. The results deviate significantly for more oblique impacts. In Eq. (\ref{Dc_uehara}), $\mu_{rep}$ is the macroscopic friction measured as the tangent of the angle of repose.

\subsection{\label{fate} Projectile fate}

\begin{figure}[ht]
	\begin{center}
		\includegraphics[width=\linewidth]{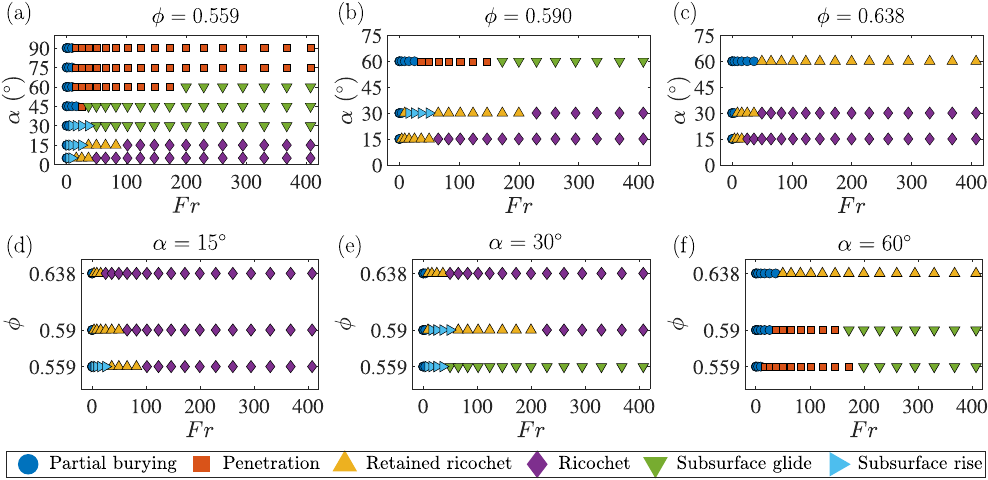}\\
	\end{center}
	\caption{Classification maps of the projectile fate for non-spinning projectiles. (a)-(c) Maps in the $\alpha$ - Fr space for different packing fractions: $\phi$ $=$ 0.559, 0.590 and 0.638, respectively. (d)-(f) Maps in the $\phi$ - Fr space for different impacting angles: $\alpha$ $=$ 15$^{\circ}$, 30$^{\circ}$ and 60$^{\circ}$, respectively. The symbols are listed in the key.}
	\label{fig:fate2}
\end{figure}

\begin{figure}[ht]
	\begin{center}
		\includegraphics[width=\linewidth]{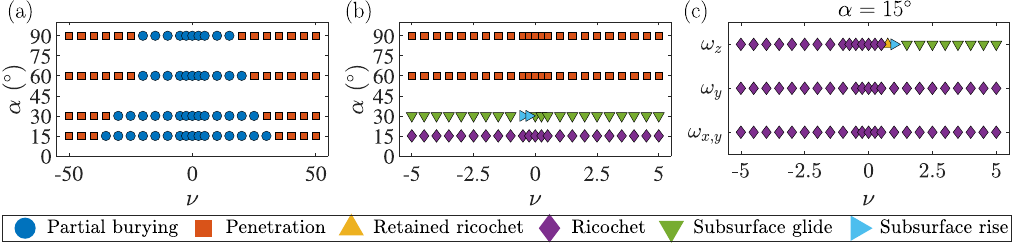}\\
	\end{center}
	\caption{Classification maps of the projectile fate for spinning projectiles. (a) and (b) Maps in the $\alpha$ - $\nu$ space for spin aligned in the impact direction ($\omega_{x,y}$) and  different impacting velocities: $V_p$ $=$ 0.5 and 5 m/s, respectively, corresponding to Fr $=$ 1 and 102. (c) Map in the $\omega$ - $\nu$ space, for $\alpha$ $=$ 15$^{\circ}$, $V_p$ $=$ 5 m/s (Fr $=$ 102), and different spin directions: vertical ($\omega_y$), transverse ($\omega_z$), and aligned in the impact direction ($\omega_{x,y}$). The symbols are listed in the key.}
	\label{fig:fate3}
\end{figure}

We now investigate how the fate of the projectile varies with $\alpha$, $V_p$, $\phi$, and $\vec{\omega}$. We begin with the non-spinning case for random loose beds ($\phi$ $=$ 0.559) shown in Fig. \ref{fig:fate2}(a) in the $\alpha$ - Fr map. For $\alpha$ $\gtrsim$ 70$^{\circ}$, this figure shows that partial burying takes place for Fr $\lesssim$ 10 (small velocities) while penetration occurs for Fr $\gtrsim$ 10 (higher velocities). When 40$^{\circ}$ $\lesssim$ $\alpha$ $\lesssim$ 70$^{\circ}$, partial burying continues for low Fr, and penetration occurs for intermediate Fr numbers (within 10 and 200, depending on the incident angle), with its range in terms of Fr decreasing as $\alpha$ decreases. For larger values of Fr (from within 10 and 200 on, depending on the incident angle), subsurface glide takes place. As the value of $\alpha$ decreases, subsurface rise replaces partial burying for intermediate values of Fr (Fr $\lesssim$ 50) when $\alpha$ $=$ 30$^{\circ}$, with subsurface glide still present for Fr $\gtrsim$ 50. When $\alpha$ $<$ 30$^{\circ}$, subsurface glide is replaced by retained ricochet (Fr $\lesssim$ 100) followed by ricochet. When we increase the packing fraction to $\phi$ = 0.590 and 0.638, Figures \ref{fig:fate2}(b) and \ref{fig:fate2}(c) show that the behavior in the $\alpha$ - Fr space is similar, with thresholds for $\alpha$ and Fr changing, and penetration and subsurface glide disappearing for $\phi$ $=$ 0.638. These observations are corroborated by the $\phi$ - Fr maps (Figs. \ref{fig:fate2}(d)-(f)). Our results follow the trend reported in the literature. For instance, our data for low angles agree reasonable well with Figs. 8 and 15 of Wright et al. \cite{Wright}, any discrepancies in thresholds being due to uncertainties in their experiments (acknowledged by them) and differences in $\phi$ (whose values are not even reported in their paper). We note that, different from previous works, we systematically varied the packing fraction $\phi$ and the projectile spin $\vec{\omega}$.

Besides the packing fraction $\phi$, we expect that the projectile behavior depends strongly on $\vec{\omega}$. When $\alpha$ is low, ricochets occur if the lift force overcomes the projectile weight just after the impact takes place, meaning that the reduction due to drag of the horizontal velocity $V_p\cos(\alpha)$ of the projectile upon the impact is not enough to stop the rebound \cite{Soliman, Wright}. When $\alpha$ is higher, a spinning projectile can experience vertical and/or transverse forces due to friction with grains, those forces being called Magnus force by Kumar et al. \cite {Kumar} (although the mechanism within grains is rather different than in a fluid, and the direction is inverted with respect to fluids). For a spinning projectile with angular velocity aligned with the impact angle, Fig. \ref{fig:fate3}(a) shows that only partial burying and penetration coexist when $V_p$ $=$ 0.5 m/s (small linear kinetic energy, corresponding to larger values of $\nu$). Partial burying occurs for low values ($|\nu|$ $\lesssim$ 30) and penetration for larger ($|\nu|$ $\gtrsim$ 30) values of $\omega$, the range of the former increasing with decreasing $\alpha$. By increasing the impact velocity by one order of magnitude ($V_p$ $=$ 5 m/s, corresponding to smaller values of $\nu$), only penetration takes place when $\alpha$ $\gtrsim$ 60$^{\circ}$, only ricochet occurs when $\alpha$ $\lesssim$ 20$^{\circ}$, and subsurface glide and subsurface rise take place for 20$^{\circ}$ $\lesssim$ $\alpha$ $\lesssim$ 60$^{\circ}$, the later only for values of $\omega$ (and $\nu$) close to zero. Figure \ref{fig:fate3}(c) shows the $\omega$ - $\nu$ map for $\alpha$ $=$ 15$^{\circ}$, $V_p$ $=$ 5 m/s, and different spin directions: vertical ($\omega_y$), transverse ($\omega_z$), and aligned in the impact direction ($\omega_{x,y}$). This figure shows that negative values of the transverse component of the spin, $\omega_z$, imply ricochet, since the tangential velocity of the projectile at the impact region points rearwards (converting rotational kinetic energy into linear kinetic energy), while positive values imply retained ricochet, subsurface rise, and subsurface glide (in order of augmenting $\nu$). For the vertical and longitudinal components of the spin ($\omega_y$ and $\omega_{x,y}$, respectively), we observe only ricochet, independent of the value of $\nu$.

\begin{figure}[ht]
	\begin{center}
		\includegraphics[width=0.35\linewidth]{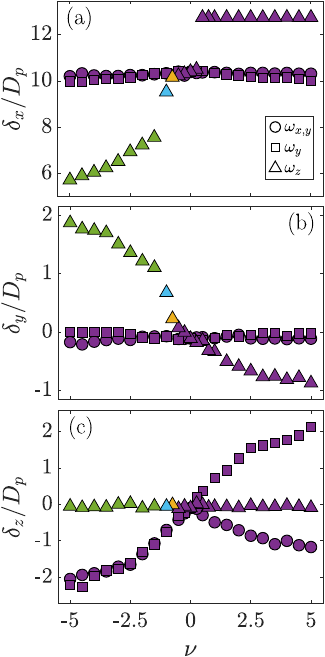}\\
	\end{center}
	\caption{(a) Distance $\delta_x$ traveled by the projectile in the $x$ direction, (b) distance $\delta_y$ in the $y$ direction, and (c) distance $\delta_z$ in the $z$ direction, as functions of $\nu$ (normalized spin). The distances are normalized by $D_p$ and the graphics are parameterized by the components of the applied spin $\vec{\omega}$: $\omega_{y}$ in the $y$ direction, $\omega_{z}$ in the $z$ direction, and $\omega_{x,y}$ in the impact direction. The symbols are listed in the key of panel (a), and the colors correspond to the projectile fate: purple corresponds to ricochet, green to subsurface glide, blue to subsurface rise, and orange to retained ricochet.}
	\label{fig:distances}
\end{figure}

\begin{figure}[ht]
	\begin{center}
		\includegraphics[width=0.35\linewidth]{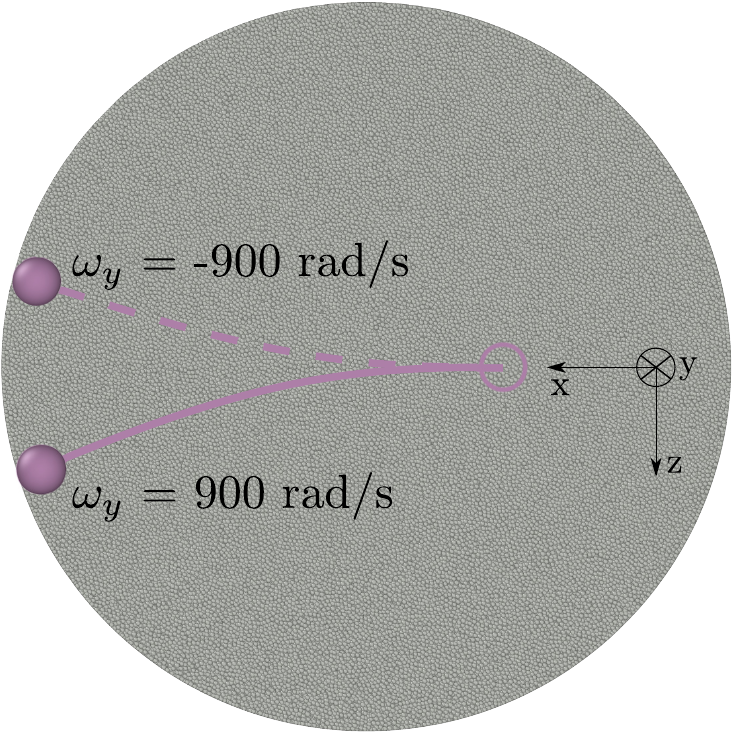}\\
	\end{center}
	\caption{Top view of the system showing the trajectories of the projectile with $\omega_{y}$ $=$ 900 rad/s (continuous line) and $\omega_{y}$ $=$ -900 rad/s (dashed line). The impact point is shown as \corr{an empty} circle and the final \corr{positions as solid} circles. In this figure, $\phi$ $=$ 0.559, $\alpha$ $=$ 15$^\circ$, and $V_p$ $=$ 5 m/s}
	\label{fig:magnus}
\end{figure}

Figures \ref{fig:distances}(a)-(c) show the $\delta_x$, $\delta_y$ and $\delta_z$ distances traveled by the projectile after the impact has taken place, in the $x$, $y$, and $z$ directions, respectively, as functions of $\nu$, for $\alpha$ $=$ 15$^{\circ}$ (smaller angle simulated), $\phi$ = 0.559 (random loose bed), and different components of the applied spin $\vec{\omega}$. The distances are normalized by the projectile diameter $D_p$. Figures \ref{fig:distances}(a) and \ref{fig:distances}(b) show that $\omega_z$ $>$ 0 promotes forward motion (and then ricochet), with $\delta_x/D_p$ saturating at 13 because the projectile leaves the domain after rebounding, and $\delta_y$ decreasing as $\omega_z$ increases. This occurs because when $\omega_z$ $>$ 0, the tangential velocity at the projectile-bed contact points backward, enhancing a forward (and upward) motion. On the contrary, when $\omega_z$ $<$ 0, the tangential velocity at the contact points forward, hindering forward motion and promoting penetration of the projectile, as can be seen in Figs. \ref{fig:distances}(a) and \ref{fig:distances}(b) (in order of increasing magnitude of $\nu$, retained ricochet, subsurface rise, and subsurface glide). Figure \ref{fig:distances}(c) shows the effect of $\omega_y$ (vertical spin) on the motion of the projectile, with $\delta_z$ $>$ 0 when $\omega_y$ $>$ 0, and $\delta_z$ $<$ 0 when $\omega_y$ $<$ 0. As in the case of $\omega_z$, the mechanism for the deviation is related to the friction and shocks at the contact point: upon and after the impact, the projectile compresses the bed just in front of it and decompresses the bed just behind it, so that there is a larger projectile-bed friction on the projectile front (with respect to its motion). Therefore, lateral motion is oriented contrary to the tangential velocity at the frontal projectile-bed contact, being to the left when $\omega_y$ points downward (and vice-versa). The force leading to both the lateral displacement in the case of $\omega_y$ and the rebound and burying enhancements in the case of $\omega_z$ was described by Kumar et al. \cite{Kumar} as the Magnus effect in granular media, although the mechanism and orientation are different from that in fluids. The origin of this Magnus force in grains is due to friction with a larger number of contact points on one side of the projectile (front), while in fluids it is due to a pressure distribution generated by the velocity field. Examples of trajectories of projectiles with positive and negative initial spins in the $y$ direction are shown in Fig. \ref{fig:magnus}. Finally, when the spin is aligned with the impact velocity, the transverse displacements are non-monotonic, $\delta_z$ acquiring always negative values.

\subsection{\label{correlation} Correlation between the projectile fate and crater morphology}

\begin{figure}[ht]
	\begin{center}
		\includegraphics[width=0.7\linewidth]{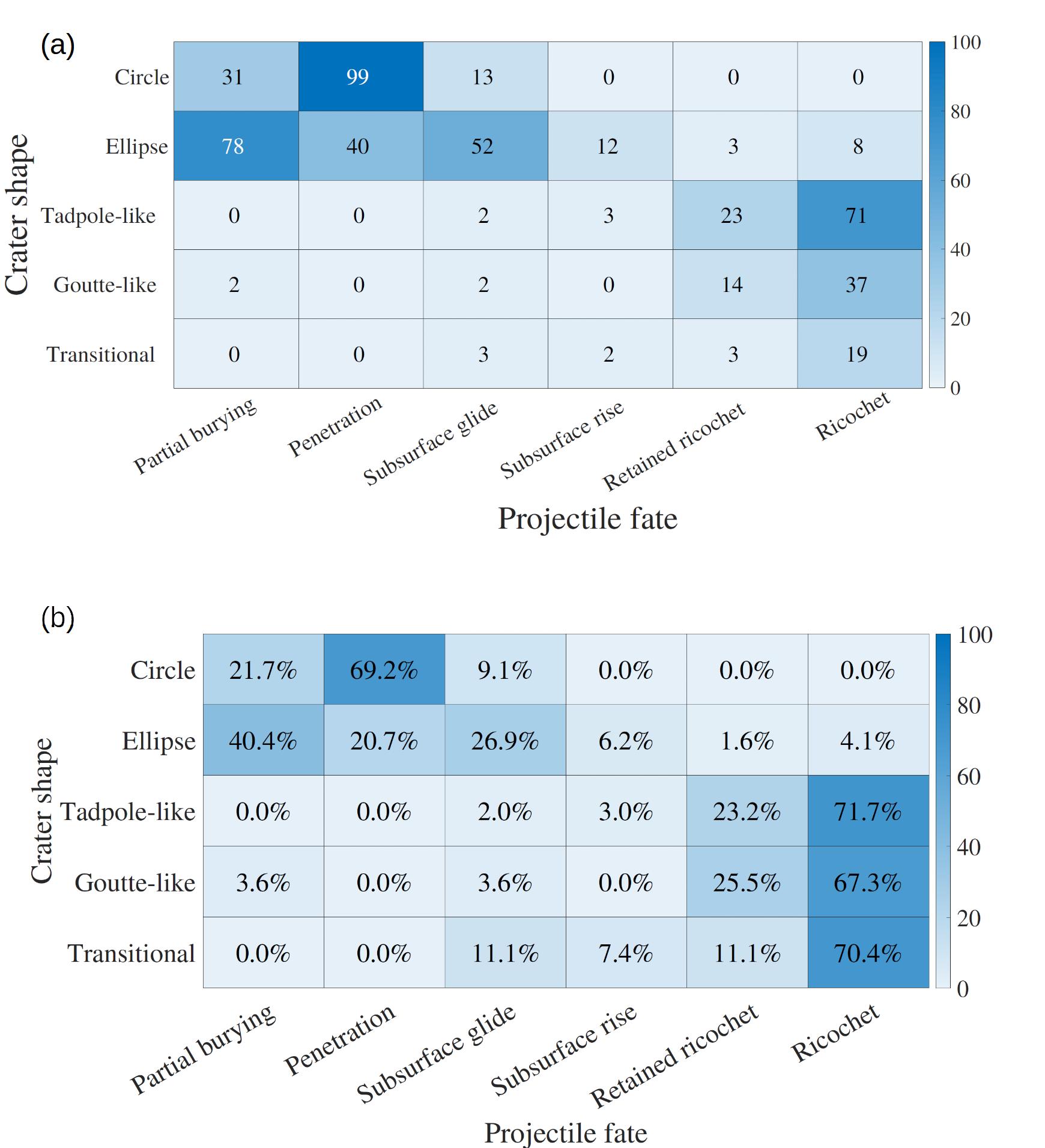}\\
	\end{center}
	\caption{\corr{(a) Diagram crossing the frequencies of occurrence crater shapes (ordinate) and projectile fates (abscissa), for all simulations. (b) Percentages of occurrence for each crater shape, as a function of the different fates. The variations of all parameters are not distinguished in the diagrams.}}
	\label{fig:correlation}
\end{figure}

Finally, we investigate if there is a correlation between the fate of projectiles and the morphology of craters. Figure \ref{fig:correlation}(a) presents a diagram of frequencies of occurrence of both crater shapes and projectile fates, for all simulations (the variations of parameters are not distinguished). It shows that projectiles forming circular and elliptical craters undergo partial burying, penetration, and subsurface glide, with those forming circles undergoing mostly penetration while those forming ellipses undergo mostly partial burying and subsurface glide. In the cases of tadpole- and goutte-like craters, projectiles undergo mostly ricochet and retained ricochet, and for transitional craters they undergo mostly ricochet. \corr{The correlation shown in Fig. \ref{fig:correlation}(a) agrees with what we expect from the mechanics of crater formation: (i) in the penetration case, an avalanche will occur after collision, covering entirely the projectile \cite{Carvalho2, Carvalho3} (forming then a circular crater); (ii) for either partial burying, subsurface glide, or subsurface rise occurring under oblique impacts, the projectile will displace grains in an asymmetric way and form elliptical craters; (iii) when the incident angle is low and the projectile experiences either retained ricochet or ricochet, the asymmetries generated are higher than that for the elliptical crater so that tadpole-like, goutte-like, or transitional craters appear.} 

Therefore, a correlation relating the shape of crates and fate of the projectile exists, so that the final location of projectiles can be inferred from the crater shape. Figure \ref{fig:correlation}(b) shows the same diagram in terms of percentages of occurrence for each crater shape, which corroborates the previous observations. \corr{The same tendencies are also observed if we consider the rotating cases separately (please see the diagram available in the Supplemental Material \cite{Supplemental})}.

\section{CONCLUSIONS}
\label{sec:conclusions}

We carried out DEM simulations of the oblique impact of a spinning projectile onto a granular bed, where, different from previous works on oblique impacts, we varied the impact velocity $V_p$, the impact angle $\alpha$, the bed packing fraction $\phi$, and the initial spin of the projectile $\vec{\omega}$. By varying these parameters, we obtained five different shapes of craters, and showed under which conditions they appear: (i) circle, for $\alpha$ $\gtrsim$ 60$^{\circ}$, and medium- to loose-packed beds ($\phi$ $\lesssim$ 0.590); (ii) ellipse, for $\alpha$ $\lesssim$ 60$^{\circ}$, being more frequent for medium- to loose-packed beds ($\phi$ $\lesssim$ 0.590); (iii) tadpole-like shape, appearing when $\alpha$ $\lesssim$ 30$^{\circ}$ at all values of $\phi$ and moderate values of $V_p$; (iv) goutte-like shape, appearing mostly for non-spinning projectiles impacting medium- to high-packed beds ($\phi$ $\gtrsim$ 0.590); and (v) transitional shape (shape between those in iii and iv), coexisting with the tadpole- and goutte-like shapes in the regions where the latter are found.

We also found the fates of the projectile, and correlated them with the crater morphology. Our results show that, for non-spinning projectiles, partial burying occurs for $\alpha$ $\gtrsim$ 70$^{\circ}$ and low Fr, while penetration takes place for $\alpha$ $\gtrsim$ 70$^{\circ}$, high Fr, and medium- to loose-packed beds ($\phi$ $\lesssim$ 0.590). 
In general, subsurface glide occurs for 40$^{\circ}$ $\lesssim$ $\alpha$ $\lesssim$ 70$^{\circ}$, high Fr, and $\phi$ $\lesssim$ 0.590, subsurface rise appears for intermediate values of Fr when $\alpha$ $=$ 30$^{\circ}$ and $\phi$ $\lesssim$ 0.590, retained ricochet occurs for $\alpha$ $<$ 30$^{\circ}$  and moderate Fr, and ricochet occurs for $\alpha$ $<$ 30$^{\circ}$ and higher values of Fr.
For spinning projectiles, both the intensity and direction of the spin strongly influence the projectile fate. When aligned with the impact direction, strong spins lead to burying and partial burying, while, under low $\alpha$, transverse spins with $\omega_z$ $<$ 0 imply a tangential velocity of the projectile that points rearwards (converting, thus, rotational energy into linear kinetic energy), leading to ricochet. When $\omega_z$ $>$ 0, the tangential velocity points forward, hindering linear translation and leading to retained ricochet, subsurface rise or subsurface glide.
We also observed that there is a transverse motion led by the projectile rotation, and oriented contrary to the tangential velocity at the projectile-bed contact at the projectile front. Therefore, the projectile moves to the left when $\omega_y$ points downward (and vice-versa), and downwards when $\omega_z$ $<$ 0 (and vice-versa). Although called Magnus effect in granular media \cite{Kumar}, the orientation is inverted with respect to fluids, and the mechanism is frictional (not due to pressure).

Finally, we show that projectiles forming circular craters undergo mostly penetration, those forming ellipses undergo mostly partial burying and subsurface glide, those forming tadpole- and goutte-like craters undergo mostly ricochet and retained ricochet, and those forming transitional craters undergo mostly ricochet. Our findings bring new explanations for different types of craters observed in nature and the final location of the impacting material.

\section{\label{sec:Ack} ACKNOWLEDGMENTS}

The authors are grateful to the S\~ao Paulo Research Foundation -- FAPESP (Grants No. 2018/14981-7 and No. 2024/13981-4) and Conselho Nacional de Desenvolvimento Cient\'ifico e Tecnol\'ogico -- CNPq (Grant No. 405512/2022-8) for the financial support provided.

\bibliography{references}

\end{document}